\documentclass[12pt]{iopart}

\eqnobysec
\usepackage{graphicx}
\usepackage{iopams}
\usepackage{color}
\usepackage{subfigure}

\begin{document}

\title[Non-conventional Anderson localization in a matched quarter stack]
{Non-conventional Anderson localization in a matched quarter stack with metamaterials}

\author{E ~J ~Torres-Herrera${}^{1}$, F ~M ~Izrailev${}^{1}$ and N~M~Makarov${}^{2}$}

\address{${}^{1}$ Instituto de F\'{\i}sica, Universidad Aut\'{o}noma de Puebla, Apartado Postal J-48, Puebla, Pue., 72570, M\'{e}xico}

\address{${}^{2}$ Instituto de Ciencias, Universidad Aut\'{o}noma de Puebla, Priv. 17 Norte No~3417, Col. San Miguel Hueyotlipan, Puebla, Pue., 72050, M\'{e}xico}

\ead{felix.izrailev@gmail.com and makarov.n@gmail.com}

\begin{abstract}
We study the problem of non-conventional Anderson localization emerging in bilayer periodic-on-average structures with alternating layers of materials with positive and negative refraction indices $n_a$ and $n_b$. Main attention is paid to the model of the so-called quarter stack with perfectly matched layers (the same unperturbed by disorder impedances, $Z_a=Z_b$, and optical path lengths, $n_ad_a=|n_b| d_b$, with $d_a$, $d_b$ being the thicknesses of basic layers). As was recently numerically discovered, in such structures with weak fluctuations of refractive indices (compositional disorder) the localization length $L_{loc}$ is enormously large in comparison with the conventional localization occurring in the structures with positive refraction indices only. In this paper we develop a new approach which allows us to derive the expression for $L_{loc}$ for weak disorder and any wave frequency $\omega$. In the limit $\omega \rightarrow 0$ one gets a quite specific dependence, $L^{-1}_{loc}\propto\sigma^4\omega^8$ which is obtained within the fourth order of perturbation theory. We also analyze the interplay between two types of disorder, when in addition to the fluctuations of $n_a$, $n_b$ the thicknesses $d_a$, $d_b$ slightly fluctuate as well (positional disorder). We show how the conventional localization recovers with an addition of positional disorder.
\end{abstract}



\submitto{\NJP}

\maketitle

\section{Introduction}
\label{sec:Intro}

Due to a remarkable progress in manufacturing one-dimensional (1D) systems with given transport characteristics the interest to a rigorous analysis of various 1D models has greatly increased in recent decades (see, e.g., book~\cite{MS08} and references therein). Of a particular interest are {\it bilayer structures} in optics \cite{optics} and electromagnetics \cite{electromagn}, semiconductor superlattices \cite{superlattice}, in the devices with alternating quantum wells and barriers in electronics, etc.

Unlike the disordered models with continuous potentials for which the theory is fully developed, the analysis of the Kroing-Penney type models meets serious theoretical difficulties. One of the first non-trivial problems was a famous effect of band center anomaly in the standard tight-binding Anderson model. As was found numerically \cite{CKM81}, the value of the localization length at the band center did not support simple analytical predictions based on the conventional perturbation theory \cite{T79}. More careful analysis performed in \cite{KW81,IRT98} have shown that one has to use specific technics in order to derive correct results. The origin of the discovered effect was found to be related to a kind of resonances emerging for weak disorder. As was later understood, such an anomaly is due to a non-homogeneous distribution of the phase of a wave function, generated in the process of wave propagation along a structure. Similar effects are also known to occur in the vicinity of band edges of energy spectra (see, for example, \cite{IKM12} and references therein).

To date, it is understood that the effect of resonance transmission of a wave through any periodic structure poses a similar problem of a correct description of various transport characteristics for the models that are {\it periodic-on-average}. The latter term refers to various kinds of weak disorder added to a strictly periodic potential. This type of systems emerge naturally since in practice on the top of periodic structures such as photonic lattices or electron superlattices, weak random variations of parameters are experimentally unavoidable. Such variations can occur for the width of barriers or wells, electromagnetic or optical characteristics of materials, effective masses of electrons in superlattices, etc. For the frequency of waves far from the resonances the phase distribution of propagating waves is typically constant. In this case the analytical treatment of the transmission coefficient or localization length is relatively easy. However, in the vicinity of band edges or resonances the distribution of wave phase turns out to be highly non-homogeneous, the fact that leads to a dramatic complication of a rigorous analysis (see review \cite{IKM12}).

Another effect which needs specific analytical tools in its analysis is the influence of various correlations in a disorder. As is already understood, the underlying correlations, either of short or long-range type, can significantly suppress or enhance the localization length, therefore, strongly influence the global characteristics of the transmission \cite{IKM12}. In particular, by imposing specific long-range correlations one can practically create the devices with given transmission/reflection characteristics. This effect of enhancement or suppression of the Anderson localization was recently observed experimentally in the waveguides with both the bulk \cite{Ulle1} or surface \cite{Ulle2} correlated disorder. It should be stressed that specific long-range correlations can emerge naturally in physical systems, and not only in the systems with intentionally included correlations. One of such examples will be discussed in this paper.

New perspectives in creating the devices with unusual transport characteristics are related to specific optic properties of metamaterials embedded in periodic structures \cite{SSS01,Go06,Ao07,Ao10,Ao10a,Mo10,MOMP12,BO12,Ao12,RBCO12,Go12}. A particular system of an increasing interest is an array of two alternating $a\,$- and $b\,$-layers with equal optical path lengths, $n_ad_a=|n_b|d_b$. Here the $a\,$-layers are made of right-handed (RH) material with positive refractive index, $n_a>0$ and thickness $d_a$. In contrast, the $b\,$-layers refer to the left-handed (LH) material with negative index, $n_b<0$, and thickness $d_b$. In such a model the phase shift of the wave gained in the $a\,$-layer is fully compensated by the subsequent shift in the next $b\,$-layer. As a result, the total phase shift vanishes after passing any of $N$ units constructed by ($a,b$) cells. If, in addition, the layers are matched ($Z_a=Z_b$), this results in a kind of invisibility of the structure for an observer.

Recent numerical data~\cite{Ao07} obtained for the array of two matched alternating RH and LH layers with $d_a=d_b$ and weakly disordered refractive indices $n_a\approx |n_b|\approx1$, have demonstrated enormously fast divergence of the localization length, $L_{loc}^{-1}\propto\omega^{6}$ for $\omega\to0$. More detailed numerical study~\cite{Ao10} has shown that the power $\kappa$ in dependence $L_{loc}^{-1}\propto\omega^{\kappa}$ increases with the sample length, and approaches the value $\kappa\approx8.78$. This result is in contrast with the conventional dependence $L_{loc}^{-1}\propto\omega^2$  known to occur for many models, both with continuous and periodic-on-average potentials.

As was found in article~\cite{TIM11}, for the model discussed in~\cite{Ao07,Ao10} the phase of wave propagating through the RH-LH array with fluctuating refractive indices is described by a highly non-uniform distribution. This effect is similar to that arising in the tight-binding Anderson model for the energy close to band edges. Further analysis \cite{TIM11} has led to a remarkable conclusion that the localization length diverges in the quadratic approximation in disorder strength. In the next study \cite{TIM12} a new method was suggested allowing one to resolve the above problem of non-conventional Anderson localization. It was shown that the inverse localization length within the fourth-order perturbation theory is described by the asymptotics $L_{loc}^{-1}\propto\sigma^{4}\omega^{8}$ for small $\omega$. Therefore, the found dependencies with $\kappa\approx6$ have to be attributed to a not large enough size $N$ of samples used in numerical calculations.

The aim of this paper is two-fold. First, we present a full description of the method used in \cite{TIM12} when analyzing the abnormal localization, numerically observed in \cite{Ao07,Ao10}. We generalize the model studied in \cite{Ao07,Ao10,TIM11,TIM12} and give many details which are important for understanding our method, together with a comprehensive discussion of the obtained results. Second, we present new analytical results for the model with two kinds of disorder, namely, with the positional and compositional disorders, and show the interplay between two mechanisms of localization. We show how the abnormal localization reduces to the conventional one in dependence on the relative strengths of the two disorders.

\section{Model formulation: Hamiltonian map}
\label{sec:ModForm}

We consider the propagation of an electromagnetic wave of frequency $\omega$ in an infinite dielectric array (stack) of two alternating $a\,$- and $b\,$-layers. Every kind of layers is respectively specified by their thickness $d_{a,b}$, dielectric permittivity $\varepsilon_{a,b}$, magnetic permeability $\mu_{a,b}$, refractive index $n_{a,b}=\sqrt{\varepsilon_{a,b}\mu_{a,b}}$, impedance $Z_{a,b}=\mu_{a,b}/n_{a,b}$ and wave number $k_{a,b}=\omega n_{a,b}/c$. We shall analyze two systems: the \emph{homogeneous} stack in which both $a\,$- and $b\,$-layers are made of RH materials, and \emph{mixed} stack where the $a\,$-layers contain the RH material while $b\,$-layers are composed of the left-handed (LH) material. As is known, in the RH medium all optic parameters are positive. On the contrary, in the LH medium the permittivity, permeability and, consequently, the refractive index are negative, however, the impedance remains to be positive. In what follows, we consider the situation when the disorder is originated from random variations of both layer thicknesses $d_{a,b}$ (\emph{positional disorder}), as well as from the fluctuations of
$\varepsilon_{a,b}$ (\emph{compositional disorder}). Specifically, we assume a weakness of both types of disorder:
\numparts
\begin{eqnarray}
d_{an}=d_a[1+\varrho_{a}(n)],\qquad\langle d_{an}\rangle=d_a\,;\label{eq:da-n}\\[6pt]
d_{bn}=d_b[1+\varrho_{b}(n)],\qquad\langle d_{bn}\rangle=d_b\,.\label{eq:db-n}
\end{eqnarray}
\endnumparts
and
\numparts
\begin{eqnarray}
\varepsilon_{an}=\varepsilon_a[1+\eta_{a}(n)]^2,\qquad n_{an}=n_a[1+\eta_{a}(n)],\nonumber\\[6pt]
Z_{an}=Z_a[1+\eta_{a}(n)]^{-1},\qquad k_{an}=\omega n_a[1+\eta_{a}(n)]/c\,; \label{eq:ka}\\[6pt]
\varepsilon_{bn}=\varepsilon_b[1+\eta_{b}(n)]^2,\qquad n_{bn}=n_b[1+\eta_{b}(n)],\nonumber\\[6pt]
Z_{bn}=Z_b[1+\eta_{b}(n)]^{-1},\qquad k_{bn}=\omega n_b[1+\eta_{b}(n)]/c\,.\label{eq:kb}
\end{eqnarray}
\endnumparts
Here the index $n$ enumerates the $n$-th unit $(a,b)$ cell, and we assume that the magnetic permeabilities $\mu_{a}$ and $\mu_{b}$ are disorder-independent.

The random sequences $\varrho_{a,b}(n)$ and $\eta_{a,b}(n)$ imposing, respectively, the positional and compositional disorder are specified by white-noise entries with the zero average and small variances $\sigma^2_\varrho,\sigma^2_\eta \ll 1$,
\begin{eqnarray}\label{eq:RhoEta-def}
\langle\varrho_{a,b}(n)\rangle=0,\qquad\langle\varrho_{a,b}^2(n)\rangle=\sigma_\varrho^2,\qquad \langle\varrho_{a}(n)\varrho_{b}(n')\rangle=\sigma_\varrho^2\delta_{ab}\delta_{nn'}\,;\nonumber\\[6pt]
\langle\eta_{a,b}(n)\rangle=0,\qquad\langle\eta_{a,b}^2(n)\rangle=\sigma_\eta^2,\qquad \langle\eta_{a}(n)\eta_{b}(n')\rangle=\sigma_\eta^2\delta_{ab}\delta_{nn'}\,;\\[6pt]
\langle\varrho_{a,b}(n)\eta_{a,b}(n')\rangle=0\,.\nonumber
\end{eqnarray}
Hereinafter, the angular brackets $\langle\ldots\rangle$ stand for the averaging over different realizations of a random structure (\emph{ensemble averaging}) or along its single realization (\emph{spatial averaging}), which is assumed to be equivalent due to ergodicity. Numerically, when generating random sequences $\varrho_{a,b}(n)$ and $\eta_{a,b}(n)$ we use the flat distribution on a finite interval. However, our analytical expressions are valid for any distribution with finite variance.

Within every $a\,$- or $b\,$-layer the electric field of the wave,
\begin{equation}\label{eq:Efield}
E(x,t)=E(x)\exp(-\rmi\omega t),
\end{equation}
obeys the one-dimensional Helmholtz equation with two boundary conditions at the interfaces between neighboring layers,
\begin{eqnarray}
&&\left(\frac{\rmd^2}{\rmd x^2}+k_{a,b}^2\right)E_{a,b}(x)=0,\label{eq:WaveEq-ab}\\[6pt]
&&E_a(x_i)=E_b(x_i),\qquad \mu_a^{-1}E'_a(x_i)=\mu_b^{-1}E'_b(x_i).\label{eq:BC}
\end{eqnarray}
The $x$-axis is directed along the array of bilayers with $x=x_{i}$ standing for the interface coordinate. The prime implies the derivative with respect to $x$.

The general solution of the wave equation \eref{eq:WaveEq-ab} inside the $n$-th unit $(a,b)$ cell can be presented as a superposition of two standing waves,
\numparts
\begin{eqnarray}
\fl E_{a}(x)=E_{a}(x_{a_n})\cos\left[k_{an}(x-x_{a_n})\right]+k_{an}^{-1}E'_a(x_{a_n})\sin\left[k_{an}(x-x_{a_n})\right]\label{eq:Ea}\\[6pt]
\mbox{inside}\,\,a_n\,\mbox{layer, where}\,\,x_{a_n}\leqslant x\leqslant x_{b_n}\,;\nonumber\\[6pt]
\fl E_{b}(x)=E_{b}(x_{b_n})\cos\left[k_{bn}(x-x_{b_n})\right]+k_{bn}^{-1}E'_b(x_{b_n})\sin\left[k_{bn}(x-x_{b_n})\right]\label{eq:Eb}\\[6pt]
\mbox{inside}\,\,b_n\,\mbox{layer, where}\,\,x_{b_n}\leqslant x\leqslant x_{a_{n+1}}\,.\nonumber
\end{eqnarray}
\endnumparts
The coordinates $x_{a_n}$ and $x_{b_n}$ refer to the left-hand edges of successive $a_n$ and $b_n$ layers, respectively. The thickness of individual layers is defined as
\begin{equation}\label{eq:dadb}
d_{an}=x_{b_n}-x_{a_n}\quad\mbox{and}\quad d_{bn}=x_{a_{n+1}}-x_{b_n}.
\end{equation}

The solutions \eref{eq:Ea} and \eref{eq:Eb} give rise to useful relations between the electric field $E_{a,b}$ and its derivative $E'_{a,b}$ at opposite boundaries of $a_n$ and $b_n$ layers,
\numparts
\begin{eqnarray}
E_a(x_{b_n})=E_a(x_{a_n})\cos\varphi_{an}+k_{an}^{-1}E'_a(x_{a_n})\sin\varphi_{an},\nonumber\\[6pt]
E'_a(x_{b_n})=-k_{an}E_a(x_{a_n})\sin\varphi_{an}+E'_a(x_{a_n})\cos\varphi_{an};\label{eq:EaE'a}\\[6pt]
E_b(x_{a_{n+1}})=E_b(x_{b_n})\cos\varphi_{bn}+k_{bn}^{-1}E'_b(x_{b_n})\sin\varphi_{bn},\nonumber\\[6pt]
E'_b(x_{a_{n+1}})=-k_{bn}E_b(x_{b_n})\sin\varphi_{bn}+E'_b(x_{b_n})\cos\varphi_{bn}.\label{eq:EbE'b}
\end{eqnarray}
\endnumparts
The disordered phase shifts $\varphi_{an}$ and $\varphi_{bn}$ depend on the cell index $n$ due to randomized refractive indices $n_{an}$ and $n_{bn}$, as well as via random thicknesses $d_{an}$ and $d_{bn}$ of the layers,
\numparts
\begin{eqnarray}
\varphi_{an}=k_{an}d_{an}=\varphi_{a}[1+\eta_{a}(n)][1+\varrho_{a}(n)],\label{eq:phia-n}\\[6pt]
\varphi_{bn}=k_{bn}d_{bn}=\varphi_{b}[1+\eta_{b}(n)][1+\varrho_{b}(n)].\label{eq:phib-n}
\end{eqnarray}
\endnumparts
Here the unperturbed phase shifts $\varphi_{a}$ and $\varphi_{b}$ are defined by the expressions:
\begin{equation}\label{eq:phi-ab}
\varphi_{a}=k_ad_a=\omega n_ad_a/c,\qquad\varphi_{b}=k_bd_b=\omega n_bd_b/c.
\end{equation}

Then, the combination of relations \eref{eq:EaE'a}, \eref{eq:EbE'b} with the boundary conditions \eref{eq:BC} at the interfaces $x_i=x_{b_n}$ and $x_i=x_{a_{n+1}}$, yields the recurrent relations describing the wave transfer through the $n$-th unit $(a,b)$ cell,
\begin{equation}\label{eq:mapQP}
Q_{n+1}=A_nQ_n+B_nP_n\,,\qquad P_{n+1}=-C_nQ_n+D_nP_n.
\end{equation}
Here $Q_n$ and $P_n$ refer to the normalized electric field and its derivative, respectively, taken at left-hand edge of the $n$-th unit $(a,b)$ cell,

\begin{equation}\label{eq:QP-def}
Q_n=Z_a^{-1/2}E_{a}(x_{a_n}),\qquad P_n=(c/\omega)Z_a^{1/2}E'_{a}(x_{a_n}).
\end{equation}
The normalization factors in \eref{eq:QP-def} contain the {\it unperturbed} impedance $Z_a$ of $a\,$-layers, see definitions \eref{eq:ka} and \eref{eq:kb}. The randomized factors $A_n$, $B_n$, $C_n$, $D_n$ read
\begin{equation}\label{eq:ABCDn}\eqalign{
\fl A_n=\cos\varphi_{an}\cos\varphi_{bn}-Z_{an}^{-1}Z_{bn}\sin\varphi_{an}\sin\varphi_{bn},\\
\fl B_nZ_a=Z_{an}\sin\varphi_{an}\cos\varphi_{bn}+Z_{bn}\cos\varphi_{an}\sin\varphi_{bn},\\
\fl C_nZ_a^{-1}=Z_{an}^{-1}\sin\varphi_{an}\cos\varphi_{bn}+Z_{bn}^{-1}\cos\varphi_{an}\sin\varphi_{bn},\\
\fl D_n=\cos\varphi_{an}\cos\varphi_{bn}-Z_{an}Z_{bn}^{-1}\sin\varphi_{an}\sin\varphi_{bn}.}
\end{equation}
As one can see, for a non-zero disorder the coefficients \eref{eq:ABCDn} depend on the cell index $n$ due to randomized phase shifts $\varphi_{an}, \varphi_{bn}$ and impedances $Z_{an}, Z_{bn}$. It should be stressed that the phase shifts are influenced by {\it both} the compositional and
positional disorders, however, the impedances depend on the compositional disorder only, see relations \eref{eq:da-n}, \eref{eq:db-n}, \eref{eq:ka}, \eref{eq:kb} and \eref{eq:phia-n}, \eref{eq:phib-n}. Note also that the recurrent relations \eref{eq:mapQP} belong to the class of area-preserving maps whose determinant equals one,
\begin{equation}\label{eq:Det1}
A_nD_n+B_nC_n=1.
\end{equation}

Remarkably, the relations \eref{eq:mapQP} can be treated as the \emph{Hamiltonian map} describing the evolution of trajectories in discrete time $n$ for a linear oscillator subjected to a time-dependent parametric force. In such a representation $Q_n$ and $P_n$ can be treated as the classical coordinate and momentum, respectively. Therefore, the problem of quantum localization can be formally reduced to the analysis of the properties of trajectories in the phase space $(Q,P)$. For the analytical treatment it is convenient to pass to polar coordinates, namely, to the radius $R_n$ and angle $\theta_n$,
\begin{equation}\label{eq:QP-RTheta}
Q_n=R_n\cos\theta_n,\qquad P_n=R_n\sin\theta_n.
\end{equation}
It can be shown that the Hamiltonian map in the \emph{radius-angle presentation} gets the form,
\begin{equation}\label{eq:mapRTheta}
\left(\frac{R_{n+1}}{R_n}\right)^{-2}=\frac{\rmd\theta_{n+1}}{\rmd\theta_n},\qquad \theta_{n+1}=\arctan\left(\frac{-C_n+D_n\tan\theta_n}{A_n+B_n\tan\theta_n}\right).
\end{equation}
Then the localization length $L_{loc}$ can be derived as the inverse of the Lyapunov exponent $\lambda$ in accordance with the definition \cite{IKM12,TIM11},
\begin{equation}\label{eq:Lyap-def}
\frac{d}{L_{loc}}\equiv\lambda=\frac{1}{2}\left\langle\ln\left(\frac{R_{n+1}}{R_{n}}\right)^2
\right\rangle= -\frac{1}{2}\left\langle\ln\frac{\rmd\theta_{n+1}}{\rmd\theta_{n}}\right\rangle.
\end{equation}
Here the average $\langle\ldots\rangle$ is performed along the ''time" $n$, with a consequent average over different realizations of disorder in order to reduce the fluctuations. Such a representation is very useful since instead of the two-dimensional map for $R_n$ and $\theta_n$
the analysis of the Lyapunov exponent can be done in terms of the one-dimensional map for the phase $\theta_n$. It should be, however, noted that the expression (\ref{eq:Lyap-def}) is not valid in the vicinity of band edges, where special methods have to be used (see \cite{IKM12} and
references therein).

\section{Unperturbed periodic counterpart: Matched quarter stack}
\label{sec:UnpertStack}

Without disorder ($\varrho_{a,b}=0$ and $\eta_{a,b}=0$) all unit cells are identical since the parameters of $a\,$- and $b\,$-layers do not depend on the cell index $n$. Therefore, the array of bilayers is periodic with the period $d$ which is the unperturbed size of a unit $(a,b)$ cell,
\begin{equation}\label{eq:d-def}
d=d_a+d_b.
\end{equation}
As is known, the transmission through a periodic bilayer stack-structure is governed by the following dispersion relation~\cite{MS08},
\begin{equation}\label{eq:DR-ab}
\cos\gamma=\cos\varphi_{a}\cos\varphi_{b}-\frac{1}{2}\left(\frac{Z_a}{Z_b}+\frac{Z_b}{Z_a}\right)\sin\varphi_{a}\sin\varphi_{b},
\end{equation}
which defines the Bloch wave number $\kappa\equiv\gamma/d$. We remind that $\varphi_{a}$ and $\varphi_{b}$ are the unperturbed phase shifts given by \eref{eq:phi-ab}.

The dispersion relation \eref{eq:DR-ab} specifies the band-structure in the dependence of the Bloch phase $\gamma$ on the wave frequency $\omega$. Within the \emph{spectral bands} where $|\cos\gamma|<1$, the solution $\gamma(\omega)$ is real, therefore, electromagnetic waves can propagate through the bilayer stack. Otherwise, $\gamma$ is purely imaginary inside the \emph{spectral gaps} where $|\cos\gamma|>1$; here the waves are the evanescent Bloch states attenuated on the scale of the order of $|\gamma|^{-1}$. As a result, inside spectral gaps the transmission is exponentially small for the samples with sufficiently large number $N$ of unit ($a,b$) cells, $N\gg|\gamma|^{-1}$.

In what follows, we assume that without disorder the basic $a\,$- and $b\,$-layers are \emph{perfectly matched} which means that their unperturbed impedances are equal,
\begin{equation}\label{eq:Matching}
Z_a=Z_b.
\end{equation}
In this case there are no waves reflected from the interfaces between matched layers. Therefore, in accordance with the dispersion relation \eref{eq:DR-ab}, such a stack-structure is equivalent to a homogeneous medium with the linear spectrum and mean refractive index $\overline{n}$,
\begin{equation}\label{eq:DR-Matching}
\kappa\equiv\gamma/d=\omega\overline{n}/c,\qquad \overline{n}=\frac{n_ad_a+n_bd_b}{d_a+d_b}\,.
\end{equation}
Note that for this structure there are no gaps in the spectrum \eref{eq:DR-Matching}, and the Bloch phase is simply the total phase shift of the wave after passing any unit ($a,b$) cell,
\begin{equation}\label{eq:gamma-Matching}
\gamma=\varphi_a+\varphi_b=\omega(n_ad_a+n_bd_b)/c.
\end{equation}

As we noted above, for a homogeneous RH-RH array all optic parameters are positive quantities. On the other hand, for a mixed RH-LH stack the parameters $\varepsilon_a$, $\mu_a$, and $n_a$ of $a\,$-layers are positive, whereas in $b\,$-layers $\varepsilon_b$, $\mu_b$, and $n_b$ are
negative. Therefore, the phase shift in the LH $b\,$-layers is negative, $\varphi_{b}=-\omega|n_b|d_b/c$, however, due to their definitions both impedances, $Z_a$ and $Z_b$, remain to be positive. As one can see, the only difference in the dispersion relation \eref{eq:DR-ab} is due to the sign before the second term. Specifically, for the RH-LH stack the sign ``minus" in \eref{eq:DR-ab} has to be replaced by ``plus", as compared with the RH-RH stack. Correspondingly, in the expressions \eref{eq:DR-Matching} and \eref{eq:gamma-Matching} for the RH-LH stack one has to make the substitution, $n_b\rightarrow-|n_b|$. As we show below, the change of the sign in the dispersion relation results in a strong change of properties of the localization length, therefore, in basic characteristics of the transmission.

\begin{figure}[t!]
\begin{center}
\includegraphics[scale=0.7]{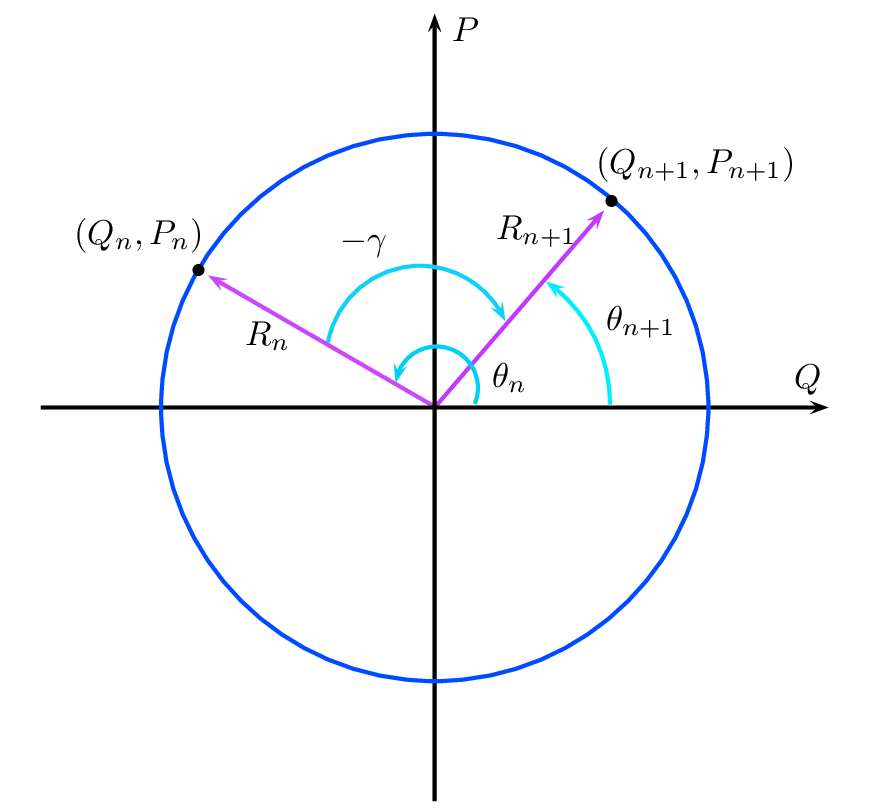}
\end{center}
\vspace{-0.5cm}
\caption{(Color online) Unperturbed Hamiltonian map \eref{eq:mapQP-unpert}, \eref{eq:mapRTheta-unpert} of periodic bilayer stack-structure.}
\label{NJP-Fig01}
\end{figure}

Let us now address the Hamiltonian map \eref{eq:mapQP} -- \eref{eq:mapRTheta}. For a periodic (without disorder) bilayer stack the coefficients \eref{eq:ABCDn} are independent of the cell index $n$. Therefore, in line with \eref{eq:Matching} and \eref{eq:gamma-Matching} the Hamiltonian map \eref{eq:mapQP} takes the following form
\begin{equation}\label{eq:mapQP-unpert}
Q_{n+1}=Q_n\cos\gamma+P_n\sin\gamma,\qquad P_{n+1}=-Q_n\sin\gamma+P_n\cos\gamma.
\end{equation}
Correspondingly, the map \eref{eq:mapRTheta} in the radius-angle presentation transforms into the relations
\begin{equation}\label{eq:mapRTheta-unpert}
R_{n+1}=R_n,\qquad \theta_{n+1}=\theta_{n}-\gamma.
\end{equation}
In spite of apparently simple forms of the unperturbed maps \eref{eq:mapQP-unpert}, \eref{eq:mapRTheta-unpert} their trajectories can be highly non-trivial. To show this, let us consider the \emph{matched quarter stack} for which two basic $a\,$- and $b\,$-layers not only have the equal impedances
\eref{eq:Matching}, but also the same lengths of optical paths,
\begin{equation}\label{eq:QStack-def}
n_ad_a=|n_b|d_b\,.
\end{equation}
For the homogeneous RH-RH array the refractive index of any $b\,$-layer is positive, $n_b>0$, therefore, the phase shifts are exactly the same, $\varphi_a=\varphi_b$, and the Bloch phase is always non-zero, $\gamma=2\varphi_a$. In this case the unperturbed map generates a circle in the
phase space $(Q,P)$ with a fixed radius $R_n$. As for the angle $\theta_n$, it changes by the Bloch phase $\gamma$ in one step of the discrete time $n$, see Figure~\ref{NJP-Fig01}.

On the contrary, for the mixed RH-LH stack, the refractive index of $b\,$-layer is negative, $n_b<0$, therefore, $\varphi_b=-\varphi_a$. This means that the phase shift $\varphi_a$ of the wave, gained in any RH layer, is canceled by the subsequent negative shift $\varphi_b$ in the next LH layer. Therefore, the phase shift $\gamma$ vanishes after passing every unit $(a,b)$ cell, $\gamma=0$. Together with the perfect transmission, this means that the structure consisting of a finite set of the RH-LH cells, has to be invisible for an observer. For such a kind of systems, the unperturbed trajectory in the phase space $(Q,P)$ degenerates into a single stationary point, the fact which is drastically different as compared with the homogeneous RH-RH stack.

Thus, when using the perturbation methods one has to take into account that in the zero order of perturbation the properties of trajectories of the Hamiltonian map remarkably depend on the type of a bilayer structure (fully conventional or having the metamaterial parts). Evidently, this fact
should be reflected by the peculiarities of the Anderson localization. In particular, one can expect that the localization length for the RH-LH structure has a non-standard dependence on the disorder, as well as on control parameters of the underlying unperturbed system.

\section{Weak compositional disorder: Phase distribution}
\label{sec:PhaseDistr}

Now we are in a position to turn on the disorder. As was shown in \cite{IKM12,IM09}, if the unperturbed impedances of two basic layers are equal, $Z_a=Z_b$, the localization length caused by the positional disorder, diverges. Indeed, the Anderson localization is originated from the effect
of multiple wave reflections. However, the impedances are independent of the disorder in the layer thicknesses $d_{an}$, $d_{bn}$, see definitions \eref{eq:da-n}, \eref{eq:db-n}, \eref{eq:ka}, \eref{eq:kb}. Consequently, the matching condition \eref{eq:Matching} remains valid even in the
presence of positional disorder. Therefore, there is no any reflection from the interfaces even if these interfaces are randomly appear in the structure due to the positional disorder. Thus, for the structure with matched layers the positional disorder has an impact on transport characteristics only being accompanied by the compositional disorder, since the latter destroys the matching of layers.

Taking into account this fact, as a first step it is reasonable to analyse the localization length contributed by the compositional disorder alone ($\varrho_{a,b}=0$). In this case all layers have constant thicknesses $d_a$ or $d_b$, so that the size $d=d_a+d_b$ of a unit ($a,b$) cell is also constant. The effect of positional disorder will be analyzed in the last part of the paper. In order to proceed, it is convenient to rewrite the definitions \eref{eq:Matching}, \eref{eq:QStack-def} for the unperturbed matched quarter stack in the form,
\begin{equation}\label{eq:MatchedQStack-def}
\fl Z_a=Z_b,\qquad\varphi_{a}=\varphi,\quad\varphi_{b}=\pm\varphi,\quad\varphi=\omega n_ad_a/c=\omega|n_b|d_b/c>0\,.
\end{equation}
Hereafter, the upper sign corresponds to the homogeneous RH-RH array, while the lower sign is associated with the mixed RH-LH stack. In order to develop a proper perturbation theory, we assume the compositional disorder to be weak,
\begin{equation}\label{eq:CD-WeakDis}
\sigma_\eta^2\ll1\qquad\mathrm{and}\qquad(\sigma_\eta\varphi)^2\ll1.
\end{equation}
In doing so, we substitute \eref{eq:MatchedQStack-def} into \eref{eq:ABCDn} with $\varrho_{a,b}=0$, and expand the coefficients $A_n$, $B_n$, $C_n$, $D_n$ up to the second order in the compositional perturbation $\eta_{a,b}(n)\ll1$. Using the uncorrelated nature \eref{eq:RhoEta-def} of the disorder, from the exact perturbed $\theta$-map \eref{eq:mapRTheta} one can obtain its quadratic approximation,
\begin{equation}\label{eq:CD-mapTheta}
\theta_{n+1}-\theta_n=-\gamma-\eta_{a}(n)U(\theta_n)\mp\eta_{b}(n)U(\theta_n-\gamma/2)-\sigma_\eta^2W(\theta_n).
\end{equation}
Here the functions $U(\theta)$ and $W(\theta)$ are defined by
\begin{equation}\label{eq:CD-UW-def}
\begin{array}{ll}
U(\theta)=\varphi+\sin\varphi\cos(2\theta-\varphi)\,,\\
W(\theta)=\varphi[\cos(2\theta-2\varphi)\pm\cos(2\theta-2\gamma)]\\
+\sin\varphi[\sin\theta\sin(\theta-\varphi)\pm\sin(\theta-\gamma/2)\sin(\theta-\varphi-\gamma/2)]\\
+\sin^2\varphi\sin(4\theta-2\varphi-\gamma)\cos\gamma\,.
\end{array}
\end{equation}

According to definition \eref{eq:Lyap-def} of the localization length, one has to know the phase distribution that emerges as a result of iterations of the Hamiltonian map. When this distribution is uniform, the average in \eref{eq:Lyap-def} can be performed relatively easy. However,
in specific cases like the standard Anderson model with the energy at the band center or close to the band edges, the analytical derivation of the phase probability density $\rho(\theta)$ is a highly non-trivial task. The same kind of problem occurs in our model with RH-LH stacks. In order to obtain the phase distribution, one has to derive the stationary Fokker-Plank equation for $\rho(\theta)$. This can be done in the way described in \cite{G04,IKT95,IRT98,HIT08}. Specifically, we rewrite the map \eref{eq:CD-mapTheta} in the continuum limit,
\begin{equation}\label{eq:CD-mapThetaContinuum}
\rmd\theta=-\gamma\rmd t-U(\theta)\eta_a(t)\rmd t\mp U(\theta-\gamma/2)\eta_b(t)\rmd t-\sigma_\eta^2W(\theta)\rmd t,
\end{equation}
where random variables $\eta_a(t)\rmd t$, $\eta_b(t)\rmd t$ are replaced, respectively, with the Wiener variables $\rmd{\cal{W}}_a=\eta_a(t)\rmd t$ and $\rmd{\cal{W}}_b=\eta_b(t)\rmd t$. As a result, expression \eref{eq:CD-mapThetaContinuum} transforms to the stochastic It\^o equation,
\begin{eqnarray}\label{eq:CD-Ito}
\rmd\theta=-U(\theta)\rmd{\cal{W}}_{a}\mp U(\theta-\gamma/2)\rmd{\cal{W}}_{b}-\left[\gamma+\sigma_{\eta}^2W(\theta)\right]\rmd t.
\end{eqnarray}
In accordance with \eref{eq:RhoEta-def}, the Wiener variables have the white-noise properties \cite{G04}:
\begin{equation}\label{eq:CD-Wiener}
\left\langle\rmd{\cal{W}}_{a,b}\right\rangle=0,\qquad
\left\langle\rmd{\cal{W}}_a\rmd{\cal{W}}_b\right\rangle=2\sigma_{\eta}^2\delta_{ab}\rmd t.
\end{equation}
Following the theory of stochastic differential equations \cite{G04}, one can readily associate the It\^o equation \eref{eq:CD-Ito} with the stationary Fokker-Plank equation for the probability density $\rho(\theta)$,
\begin{equation}\label{eq:CD-FP}
\frac{\rmd^2}{\rmd\theta^2}\left[U^2(\theta)+U^2(\theta-\gamma/2)\right]\rho(\theta)+ 2\frac{\rmd}{\rmd\theta}\left[\frac{\gamma}{\sigma_{\eta}^2}+W(\theta)\right]\rho(\theta)=0.
\end{equation}
This equation should be complemented by the condition of periodicity and by the normalization condition,
\begin{equation}\label{eq:CD-FPCond}
\rho(\theta+\pi)=\rho(\theta),\qquad\qquad\int_{0}^{\pi}\rmd\theta\rho(\theta)=1.
\end{equation}
Note that the integration in \eref{eq:CD-FPCond} is performed within the interval $\pi$ due to the periodicity of $\rho(\theta)$ with the period $\pi$. From the above equation \eref{eq:CD-FP} one can see that its solution is sensitive to whether the Bloch phase $\gamma$ is non-zero (RH-RH
array) or vanishes (RH-LH array).

\section{RH-RH matched quarter stack}
\label{sec:RH-RH stack}

In such a structure the unperturbed optic path lengths of two basic layers are equal, and the Bloch phase \eref{eq:gamma-Matching} is non-zero,
\begin{equation}\label{eq:HomQStack-def}
n_ad_a=n_bd_b\quad\to\quad\varphi_{a}=\varphi_{b}=\varphi\quad\to\quad\gamma=2\varphi=2\omega n_ad_a/c.
\end{equation}
In this case the term in the Fokker-Plank equation \eref{eq:CD-FP} containing $\gamma/\sigma_{\eta}^2$ prevails over the others for \emph{any}, even arbitrarily small, value of the phase shift $\varphi$. Indeed, under weak disorder conditions \eref{eq:CD-WeakDis} and small $\varphi\ll1$ one can get the relations $\sigma_{\eta}^2W(\theta)/\gamma\sim\sigma_{\eta}^2\ll1$ and $\sigma_{\eta}^2U^2(\theta)/\gamma\sim\sigma_{\eta}(\sigma_{\eta}\varphi)\ll1$, which allow to neglect all the terms with $W$ and $U$ in \eref{eq:CD-FP}. As a result, the phase distribution $\rho(\theta)$ is \emph{uniform} within the first order of perturbation theory,
\begin{equation}\label{eq:CD-RhoUniform}
\rho(\theta)=1/\pi.
\end{equation}

The dimensionless inverse localization length $d/L_{loc}$ is derived by differentiating the map \eref{eq:CD-mapTheta} with respect to $\theta$ and substituting the result into \eref{eq:Lyap-def}, with a further expansion of the logarithm within the quadratic approximation in disorder. Then, the subsequent averaging can be readily performed over random entries $\eta_{a,b}(n)$ and over the angle $\theta$ with the uniform distribution function \eref{eq:CD-RhoUniform}. After some algebra, one gets \cite{TIM11,TIM12},
\begin{equation}\label{eq:CD-LyapHom}
d/L_{loc}\equiv\lambda=\sigma_{\eta}^2\sin^2\varphi,\qquad\qquad\varphi=\omega n_ad_a/c.
\end{equation}
Note that in the $\theta$-map \eref{eq:CD-mapTheta} only linear terms in the compositional perturbations $\eta_{a,b}(n)$ contribute to the Lyapunov exponent, since the last quadratic term vanishes after averaging over $\theta$. The final expression  \eref{eq:CD-LyapHom} is in a
complete correspondence with the results previously obtained in the papers \cite{IMT10,TIM11,TIM12}. When the phase shift $\varphi$ is small, the result \eref{eq:CD-LyapHom} yields the asymptotics,
\begin{equation}\label{eq:CD-LyapHomOmega}
d/L_{loc}\equiv\lambda\approx\sigma_{\eta}^2\omega^2(n_ad_a/c)^2\qquad\mathrm{for}\quad\omega\ll c/n_ad_a.
\end{equation}
This gives rise to the standard quadratic $\omega$-dependence $\lambda\propto\sigma_{\eta}^2\omega^2$ in the limit $\omega\to0$.

It should be noted that the localization length $L_{loc}(\omega)$ defined by \eref{eq:CD-LyapHom}, exhibits the \emph{Fabry-Perot resonances} emerging when the phase shift $\varphi$ of the wave passing any layer is multiple to $\pi$, i.e., when the wave frequency $\omega$ meets the condition
\begin{equation}\label{eq:CD-FPres}
\omega/c=s\pi/n_ad_a\qquad s=1,2,3,\dots.
\end{equation}
At the resonances the factor $\sin\varphi$ in expression \eref{eq:CD-LyapHom} vanishes, thus resulting in the resonance divergence of the localization length $L_{loc}$ and consequently, in the \emph{suppression of the localization}. Remarkably, in contrast to the localization length $L_{loc}$, the Fabry-Perot resonances in the Lyapunov exponent $\lambda$ represent quite broad oscillations because of a smooth $\sin$-function in expression \eref{eq:CD-LyapHomOmega}.

\section{RH-LH matched quarter stack}
\label{sec:RH-LH stack}

\begin{figure}[t!!!]
\begin{center}
\includegraphics[scale=0.7]{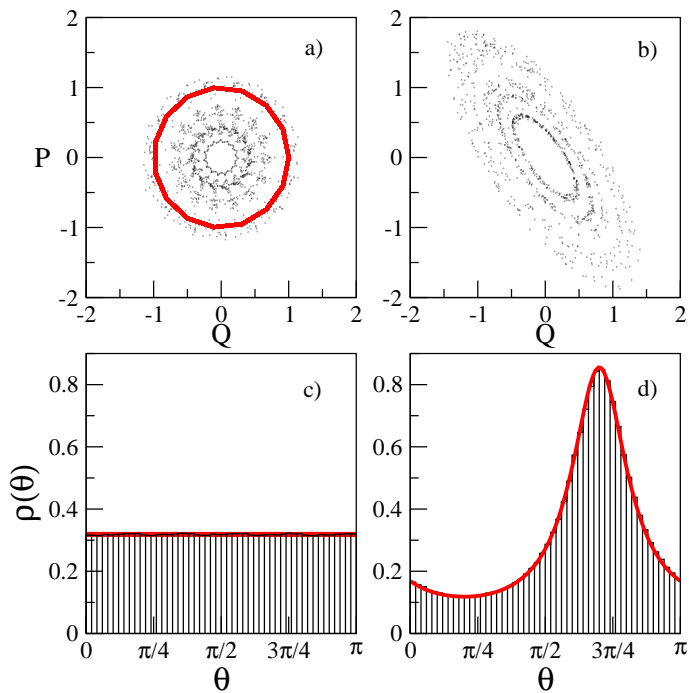}
\end{center}
\caption{(Color online) Phase trajectories generated by map \eref{eq:mapQP}: (a) RH-RH matched quarter stack with unit-cell number $N=10^4$, $\varphi=2\pi/30$, for zero disorder (solid circle), and for $\sigma_{\eta}^2=0.003$ (scattered points); (b) RH-LH matched quarter stack with $N=10^6$, $\varphi=2\pi/5$, $\sigma_{\eta}^2=0.003$. Phase distribution $\rho(\theta)$: (c) for the RH-RH array from map \eref{eq:mapQP} (histogram), and from expression \eref{eq:CD-RhoUniform} (horizontal line); (d) for the RH-LH array from map \eref{eq:mapQP} (histogram), and from expression \eref{eq:CD-RhoMix} (solid curve). Due to periodicity, $\rho(\theta+\pi)=\rho(\theta)$, only the range $0\leqslant\theta<\pi$
is shown in (c) and (d) (after \cite{TIM12}).}
\label{NJP-Fig02}
\end{figure}

The principally different situation emerges for the mixed RH-LH  matched quarter stack for which
\begin{equation}\label{eq:MixQStack-def}
n_ad_a=-n_bd_b\quad\to\quad\varphi_{a}=-\varphi_{b}=\varphi\quad\to\quad\gamma=0.
\end{equation}
In this case the Bloch phase vanishes \emph{independently} of the value of the phase shift $\varphi$. As a result, the functions \eref{eq:CD-UW-def} turn out to be related to each other, $W(\theta)=-U(\theta)U'(\theta)$, and the Fokker-Plank equation \eref{eq:CD-FP} reduces to
\begin{equation}\label{eq:CD-FPMix}
\frac{\rmd}{\rmd\theta}U(\theta)\frac{\rmd}{\rmd\theta}U(\theta)\rho(\theta)=0.
\end{equation}
Solving this equation with the conditions \eref{eq:CD-FPCond} of periodicity and normalization, we get a highly \emph{nonuniform} phase distribution,
\begin{equation}\label{eq:CD-RhoMix}
\rho(\theta)=\frac{1}{\pi}\sqrt{\varphi^2-\sin^2\varphi}\Big/U(\theta).
\end{equation}
Figure~\ref{NJP-Fig02} displays a perfect agreement between analytical expressions \eref{eq:CD-RhoUniform}, \eref{eq:CD-RhoMix} for the probability density $\rho(\theta)$ and the corresponding numerical data obtained by the direct iteration of the exact map \eref{eq:mapQP}.

The Lyapunov exponent can be derived according to its definition \eref{eq:Lyap-def} and quadratic approximation \eref{eq:CD-mapTheta} of the Hamiltonian $\theta$-map, taking into account the conditions \eref{eq:MixQStack-def}. In the second order of the perturbation theory in the weak compositional disorder $\eta_{a,b}(n)$ one gets
\begin{equation}\label{eq:CD-LyapMix2Ord}
d/L_{loc}\equiv\lambda=2\sigma_{\eta}^2\sin\varphi\langle\cos(2\theta-\varphi)U(\theta)\rangle=0.
\end{equation}
The averaging in \eref{eq:CD-LyapMix2Ord} is performed over the angle $\theta$ with the distribution function $\rho(\theta)$ determined by \eref{eq:CD-RhoMix}. Since the denominator $U(\theta)$ in \eref{eq:CD-RhoMix} is the same as the coefficient in \eref{eq:CD-LyapMix2Ord}, we come to a very unexpected result: the Lyapunov exponent vanishes within the second order perturbation theory for {\it any value} of the phase shift $\varphi$ \cite{TIM11}. This means that in order to derive a non-vanishing Lyapunov exponent, one has to go beyond the second order perturbation theory. For this, one has to obtain the expressions for both the $\theta$-map and the phase distribution $\rho(\theta)$ in the next (fourth order) approximation, which is not a simple task. The problem is that the evaluation of high order terms in $\rho(\theta)$ is not possible with the method \cite{G04,IKT95,IRT98,HIT08} applied above. The reason is that the $\theta$-map written within the fourth order approximation, cannot be reduced to the Fokker-Plank equation within the linear theory of differential stochastic equations. The only result which here can be drawn is that the Lyapunov exponent for the RH-LH matched quarter stack has to be proportional to $\sigma_{\eta}^4$, which is in contrast with the conventional quadratic dependence emerging in various disordered models.

In order to proceed further, here we suggest another method that turns out to be very effective for deriving the expression for the Lyapunov exponent. The numerical data in Figure~\ref{NJP-Fig02} give a hint of what can be done analytically. Specifically, for the RH-RH stack the trajectory in the phase space shown in Figure~\ref{NJP-Fig02}a, looks like it is created by the combination of two processes. The first one is a random increase of the radius $R_n$ defined by \eref{eq:mapRTheta}. The second process is a uniform-like filling of a circle that is confirmed by the phase distribution $\rho(\theta)$ generated by the {\it exact} Hamiltonian map. As one can see in Figure~\ref{NJP-Fig02}c, the phase distribution is uniform, the fact which has been already used for the analytical evaluation of the Lyapunov exponent. The right panel in Figure~\ref{NJP-Fig02} demonstrates completely different effect: for the RH-LH stack the points fill an ellipse characterized by some aspect ratio, and rotated by some angle $\tau$ with respect to the axes. This ellipse randomly increases in time, apparently keeping both the aspect ratio and the angle $\tau$. Thus, one can suppose that if to pass properly from the variables $Q,P$ to new ones by both rotating the axes and rescaling ellipse into the circle, the trajectory in new variables will be of the same kind as those shown in Figure~\ref{NJP-Fig02}a. In such a way one can expect that the distribution of a new phase $\widetilde{\theta}_n$ will be uniform, at least, approximately.

Following this idea, we make the linear transformation from ``old" variables $Q_n$, $P_n$ to the ``new" ones $\widetilde{Q}_n$, $\widetilde{P}_n$,
\begin{equation}\label{eq:QP-New-Old}\eqalign{
\widetilde{Q}_{n}=\alpha^{1/2}Q_n\cos\tau+\alpha^{1/2}P_n\sin\tau,\\
\widetilde{P}_{n}=-\alpha^{-1/2}Q_n\sin\tau+\alpha^{-1/2}P_n\cos\tau,}
\end{equation}
where the rotating angle $\tau$ and rescaling $\alpha$ are the parameters to be specified. Note that in the new variables the expressions \eref{eq:mapQP} and \eref{eq:QP-RTheta} -- \eref{eq:Lyap-def} conserve their forms, however, with the new factors,
\begin{equation}\label{eq:ABCDnew}\eqalign{
\widetilde{A}_n=A_n\cos^2\tau+(B_n-C_n)\sin\tau\cos\tau+D_n\sin^2\tau,\\
\widetilde{B}_n\alpha^{-1}=B_n\cos^2\tau-(A_n-D_n)\sin\tau\cos\tau+C_n\sin^2\tau,\\
\widetilde{C}_n\alpha=C_n\cos^2\tau+(A_n-D_n)\sin\tau\cos\tau+B_n\sin^2\tau,\\
\widetilde{D}_n=D_n\cos^2\tau-(B_n-C_n)\sin\tau\cos\tau+A_n\sin^2\tau,}
\end{equation}
which replace the old ones \eref{eq:ABCDn}.

Now the distribution function $\rho(\widetilde{\theta})$ for new phase $\widetilde{\theta}$ can be found starting from the quadratic expansion of the exact $\widetilde{\theta}$-map \eref{eq:mapRTheta} with new coefficients \eref{eq:ABCDnew}. Taking into account the relations \eref{eq:MixQStack-def}, we come to the following equation,
\begin{equation}\label{eq:CD-mapThetaMixNew}
\widetilde{\theta}_{n+1}-\widetilde{\theta}_{n}=[\eta_{a}(n)-\eta_{b}(n)]V(\widetilde{\theta}_n)+
\sigma_{\eta}^2V(\widetilde{\theta}_n)V'(\widetilde{\theta}_n).
\end{equation}
The stationary Fokker-Plank equation corresponding to this new $\widetilde{\theta}$-map can be obtained in the same way as described before. Then, one gets
\begin{equation}\label{eq:CD-FPMixNew}
\frac{\rmd}{\rmd\widetilde{\theta}}V(\widetilde{\theta})\frac{\rmd}{\rmd\widetilde{\theta}}V(\widetilde{\theta})\rho(\widetilde{\theta})=0.
\end{equation}
Note that the new Fokker-Plank equation \eref{eq:CD-FPMixNew} differs from the old one \eref{eq:CD-FPMix} only in the function $V(\widetilde{\theta})$ instead of $U(\theta)$. The new function $V(\widetilde{\theta})$ is defined by
\begin{eqnarray}\label{eq:CD-V-def}
&&V(\widetilde{\theta})=\sin\varphi\sin(2\tau-\varphi)\sin2\widetilde{\theta}+
\frac{\alpha}{2}[\varphi -\sin\varphi\cos(2\tau-\varphi)](\cos2\widetilde{\theta}-1)\nonumber\\
&&-\frac{\alpha^{-1}}{2}[\varphi+\sin\varphi\cos(2\tau-\varphi)](\cos2\widetilde{\theta}+1).
\end{eqnarray}
One can see that $V(\widetilde{\theta})=-U(\theta)$ at $\alpha=1$ and $\tau=0$, when new and old phases coincide.

\begin{figure}[t!!!]
\begin{center}
\includegraphics[scale=0.7]{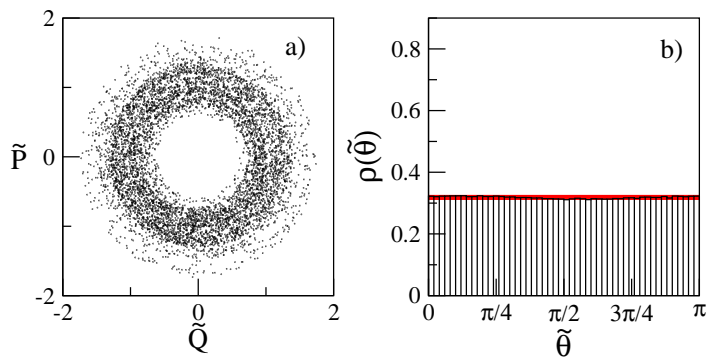}
\end{center}
\caption{(Color online) (a) Trajectory ($\widetilde{Q}_n,\widetilde{P}_n$) generated numerically by the transformed map with factors \eref{eq:ABCDnew} and parameters \eref{eq:CD-TauAlphaV}, for $\gamma=0$, $\varphi=2\pi/5$, $\sigma_{\eta}^2=0.02$ and $N=10^7$; (b) The corresponding distribution function $\rho(\widetilde{\theta})$ (after \cite{TIM12}).}
\label{NJP-Fig03}
\end{figure}

From equation \eref{eq:CD-FPMixNew} with additional conditions similar to \eref{eq:CD-FPCond}, one readily gets that the phase distribution is, indeed, uniform, $\rho(\widetilde{\theta})=1/\pi$, when the function $V(\widetilde{\theta})$ is actually independent of the angle $\widetilde{\theta}$, i.e., when its derivative with respect to $\widetilde{\theta}$ vanishes,
\begin{equation}\label{eq:CD-UniformCond}
V'(\widetilde{\theta})=0.
\end{equation}
With this condition and definition \eref{eq:CD-V-def}, we can specify the values of $\tau$ and $\alpha$ which determine the transformation from old to new phase-space variables. Also, one can obtain explicitly the $\widetilde{\theta}$-independent expression for the function $V(\widetilde{\theta})$,
\begin{equation}\label{eq:CD-TauAlphaV}
\fl \tau=\frac{\varphi}{2},\qquad\alpha^2=\frac{\varphi+\sin\varphi}{\varphi-\sin\varphi},
\qquad V(\widetilde{\theta})=V(\varphi)=\sqrt{\varphi^2-\sin^2\varphi}.
\end{equation}
The data presented in Figure~\ref{NJP-Fig03} confirm the success of our approach: in the new variables the trajectory looks like a circle with the fluctuating radius. These data correspond to our expectation that the distribution of a new phase can be considered as uniform.

The results \eref{eq:ABCDnew}, \eref{eq:CD-TauAlphaV} allow us to calculate the Lyapunov exponent $\lambda$ for the mixed RH-LH matched quarter stack with the compositional disorder. For a weak disorder \eref{eq:CD-WeakDis}, the corresponding asymptotics of the $\widetilde{\theta}$-map reads
\begin{equation}\label{eq:CD-mapThetaMix}
\widetilde{\theta}_{n+1}-\widetilde{\theta}_{n}=\left[\eta_{a}(n)-\eta_{b}(n)\right]V(\varphi)+
\frac{1}{2}\left[\eta_{a}^2(n)-\eta_{b}^2(n)\right]Y(\widetilde{\theta}_n).
\end{equation}
Here the function $Y(\widetilde{\theta})$ is defined as
\begin{equation}\label{eq:CD-Y-def}
\fl Y(\widetilde{\theta})=-\frac{(\sin\varphi-\varphi\cos\varphi)\sin\varphi}{V(\varphi)}+
\frac{\left(2 \varphi^2-\sin^2\varphi\right)\cos\varphi-\varphi\sin\varphi}{V(\varphi)}\cos2\widetilde{\theta}.
\end{equation}
Surprisingly, the fourth-order terms in the expansion of the $\widetilde{\theta}$-map do not contribute in the fourth-order approximation for the Lyapunov exponent. This happens as a result of the averaging over new phase $\widetilde{\theta}$ with the uniform distribution. Such an average is in the spirit of the perturbation theory: the next fourth-order terms can be approximated by their average over the phase, with the phase distribution obtained in the second order approximation.

Now we substitute the expression \eref{eq:CD-mapThetaMix} into definition \eref{eq:Lyap-def} for the Lyapunov exponent. Taking into account that $\lambda$ vanishes within the quadratic approximation in disorder, we expand the logarithm up to fourth-order terms in the perturbation. After, using the correlation properties \eref{eq:RhoEta-def} and averaging over $\widetilde{\theta}$ with uniform distribution function, we arrive at final result \cite{TIM12},
\begin{equation}\label{eq:CD-LyapMixed}
\frac{d}{L_{loc}}\equiv\lambda=\sigma_{\eta}^4\,\frac{\zeta+2}{4}\,
\frac{[(2\varphi^2-\sin^2\varphi)\cos\varphi-\varphi\sin\varphi]^2}{\varphi^2-\sin^2\varphi}.
\end{equation}
Here
\begin{equation}\label{eq:ExcessKurtosis}
\zeta=\frac{\langle\eta_{a,b}^4(n)\rangle}{\langle\eta_{a,b}^2(n)\rangle^2}-3
\end{equation}
is the so-called \emph{excess kurtosis}, the parameter which is specified by the form of the distribution of random variable $\eta_{a.b}(n)$. In general, $\zeta$ can have any value within the range $-2\leqslant\zeta<\infty$. In particular, for the Gaussian and flat distributions it equals $\zeta=0$ and $\zeta=-6/5$, respectively.

Expression \eref{eq:CD-LyapMixed} determines the asymptotics for small phase shift $\varphi$,
\begin{equation}\label{eq:CD-LyapMixedOmega}
\frac{d}{L_{loc}}\equiv\lambda\approx\frac{2^4}{3^35^2}(\zeta+2)\sigma_{\eta}^4\varphi^8\qquad
\mathrm{for}\quad\varphi^2\ll1\ll\sigma_{\eta}^{-2}.
\end{equation}
As one can see, a quite surprising frequency dependence of the Lyapunov exponent, $\lambda\propto\sigma_{\eta}^4\omega^8$, emerges when $\omega\to0$. Thus, the dependence $\lambda\propto\omega^6$ numerically obtained for small values of $\omega$ in Refs.~\cite{Ao07,Ao10}, can not be considered as the asymptotical result. From the analysis of the correct expression \eref{eq:CD-LyapMixed} it follows that the dependence $\lambda\propto\omega^6$ should be regarded as the intermediate one, apparently emerging due to not sufficiently large lengths $N$ over which the average of $\lambda$ was performed.

\begin{figure}[t]
\begin{center}
\includegraphics[scale=0.6]{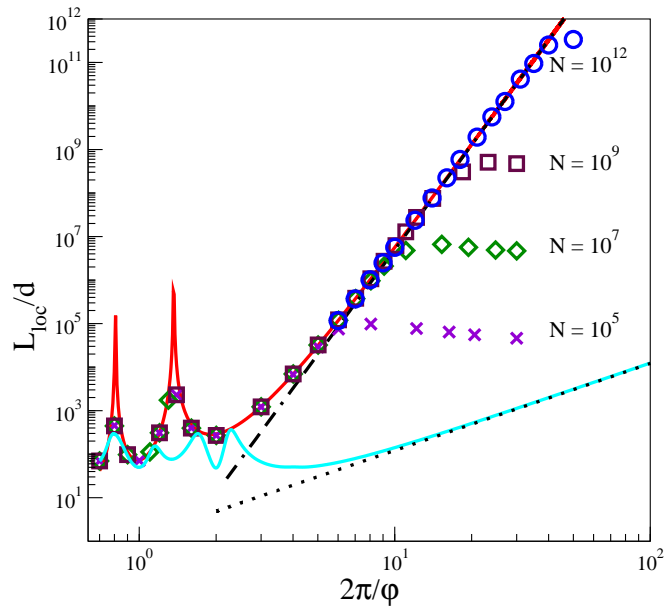}
\end{center}
\caption{(Color online) Localization length for RH-LH and RH-RH matched quarter stack of different lengths $N$ versus the normalized wavelength $2\pi/\varphi=2\pi c/n_ad_a\omega$, for $\sigma_{\eta}^2=0.02$. Full solid curve presents exact expression \eref{eq:CD-LyapMixed}, dashed-dotted line shows approximate dependence \eref{eq:CD-LyapMixedOmega}. Lowest solid curve and dotted line correspond to the expressions \eref{eq:CD-LyapHom} and \eref{eq:CD-LyapHomOmega} for RH-RH stack.}
\label{NJP-Fig04}
\end{figure}

Figure~\ref{NJP-Fig04} demonstrates an excellent agreement between the numerical data for the localization length and analytical results derived above. The data have been obtained by the iteration of the exact map \eref{eq:mapQP}, with the use of definition \eref{eq:Lyap-def}, and compared with the analytical expressions \eref{eq:CD-LyapMixed}, \eref{eq:CD-LyapMixedOmega} for the RH-LH quarter stack with $\zeta =-6/5$, as well as \eref{eq:CD-LyapHom}, \eref{eq:CD-LyapHomOmega} for the RH-RH structure. For sequences of unit-cell numbers $N=10^5, 10^7$ and $10^9$ the data are obtained with the ensemble averaging over 100 realizations of disorder, while for $N=10^{12}$ only one realization is used.

\section{RH-LH matched quarter stack: Effect of positional disorder}
\label{sec:RH-LH stack PD}

In the previous sections we presented a full analytical description of a nontrivial effect of \emph{compositional disorder} in the mixed matched quarter stack with the LH material. On the other hand, as is noted in Section~\ref{sec:PhaseDistr}, the \emph{positional disorder} itself in a bilayer stack with matched layers ($Z_a=Z_b$) does not result in the localization \cite{IKM12,IM09}. Below we show that in the presence of an additional compositional disorder, one has to take into account its non-trivial influence on the localization length. Specifically, we consider an interplay between both disorders, and show how the conventional frequency dependence of the localization length, $L^{-1}_{loc}\propto\omega^2$, (for small $\omega$) emerges due to the influence of the positional disorder.

Since we consider the mixed RH-LH matched quarter stack \eref{eq:MatchedQStack-def}, \eref{eq:MixQStack-def}, it is reasonable to treat from the beginning the \emph{transformed} Hamiltonian map, which is described by equations \eref{eq:mapQP}, \eref{eq:QP-RTheta} and \eref{eq:mapRTheta} with factors \eref{eq:ABCDnew} instead of \eref{eq:ABCDn}. The angle $\tau$ and rescaling parameter $\alpha$ are specified by expressions \eref{eq:CD-TauAlphaV}.

As before, in order to develop the perturbation theory we assume both compositional and positional disorders to be weak, compare with \eref{eq:CD-WeakDis},
\begin{equation}\label{eq:CPD-WeakDis}
\sigma_\eta^2\ll1,\qquad(\sigma_\eta\varphi)^2\ll1\qquad\mathrm{and}\qquad(\sigma_\varrho\varphi)^2\ll1.
\end{equation}
Note that the impedances \eref{eq:ka}, \eref{eq:kb}, depend only on the compositional disorder $\eta_{a,b}(n)$, whereas the phase shifts \eref{eq:phia-n}, \eref{eq:phib-n} are randomized due to both compositional $\eta_{a,b}(n)$ and positional $\varrho_{a,b}(n)$ disorders. The latter is  additionally incorporated into the layer thicknesses \eref{eq:da-n}, \eref{eq:db-n}.

The expansion of the coefficients \eref{eq:ABCDnew} and the recurrent relation \eref{eq:mapRTheta} within the quadratic approximation in the disorders $\eta_{a,b}(n)$, $\varrho_{a,b}(n)$ yields the following $\widetilde{\theta}$-map,
\begin{eqnarray}\label{eq:CPD-mapThetaMix}
\widetilde{\theta}_{n+1}-\widetilde{\theta}_{n}=\left[\eta_{a}(n)-\eta_{b}(n)\right]V(\varphi)+
[\varrho_{a}(n)-\varrho_{b}(n)]G(\widetilde{\theta}_n)\nonumber\\[6pt]
+\frac{1}{2}\left[\eta^2_a(n)-\eta^2_b(n)\right]Y(\widetilde{\theta}_n)+
\frac{1}{2}\left[\varrho_a^2(n)+\varrho_b^2(n)\right]G(\widetilde{\theta}_n)G'(\widetilde{\theta}_n).
\end{eqnarray}
This expression differs from \eref{eq:CD-mapThetaMix} in the second and fourth terms that take into account the effect of positional disorder. Functions $V(\varphi)$ and $Y(\widetilde{\theta})$ are respectively determined by \eref{eq:CD-TauAlphaV} and \eref{eq:CD-Y-def}. As for the function $G(\widetilde{\theta})$, it is given by
\begin{equation}\label{eq:CPD-G-def}
G(\widetilde{\theta})=\frac{\varphi(\varphi-\sin\varphi\cos2\widetilde{\theta})}{V(\varphi)}.
\end{equation}

The stationary Fokker-Plank equation corresponding to the $\widetilde{\theta}$-map \eref{eq:CPD-mapThetaMix} reads
\begin{equation}\label{eq:CPD-FPMix}
\sigma_\eta^2V^2(\varphi)\frac{\rmd^2}{\rmd\widetilde{\theta}^2}\rho(\widetilde{\theta})+
\sigma_\varrho^2\frac{\rmd}{\rmd\widetilde{\theta}}G(\widetilde{\theta})
\frac{\rmd}{\rmd\widetilde{\theta}}G(\widetilde{\theta})\rho(\widetilde{\theta})=0.
\end{equation}
As one can see, in this equation the first term containing the variance $\sigma_\eta^2$, is responsible for the compositional disorder, and the second one (with $\sigma_\varrho^2$) arises due to positional disorder. As both disorders do not correlate with each other, the equations
\eref{eq:CPD-mapThetaMix} and \eref{eq:CPD-FPMix} do not include the cross-correlated terms with the product $\sigma_\eta\sigma_\varrho$.

The solution to \eref{eq:CPD-FPMix} satisfying the conditions of periodicity and normalization \eref{eq:CD-FPCond} is
\begin{equation}\label{eq:CPD-RhoMix}
\rho(\widetilde{\theta})=I(\varphi)\left[\sigma_{\varrho}^2G^2(\widetilde{\theta})+\sigma_{\eta}^2V^2(\varphi)\right]^{-1/2},
\end{equation}
with the normalization function $I(\varphi)$ determined as
\begin{equation}\label{eq:CPD-I-def}
I^{-1}(\varphi)=\int_0^{\pi}\left[\sigma_{\varrho}^2G^2(\widetilde{\theta})+\sigma_{\eta}^2V^2(\varphi)
\right]^{-1/2}d\widetilde{\theta}.
\end{equation}
Remarkably, when the positional disorder vanishes the distribution function $\rho(\widetilde{\theta})$ reduces to the uniform one,
\begin{equation}\label{eq:CPD-RhoMixUn}
\rho(\widetilde{\theta})\to1/\pi\qquad\mathrm{for}\quad\sigma_{\varrho}^2\to0.
\end{equation}
This fact is in a complete accordance with the result obtained in the previous section.

The dimensionless inverse localization length $d/L_{loc}$ is obtained with the use of the map \eref{eq:CPD-mapThetaMix} in the same way as described above. Taking into account the white-noise properties \eref{eq:RhoEta-def}, the averaging over $\widetilde{\theta}$ with the probability density \eref{eq:CPD-RhoMix}, \eref{eq:CPD-I-def} yields
\begin{eqnarray}\label{eq:CPD-LyapMixed}
\frac{d}{L_{loc}}\equiv\lambda=-\sigma_{\varrho}^2\,\frac{I(\varphi)}{2}
\int_0^{\pi}G(\widetilde{\theta})G''(\widetilde{\theta})\left[\sigma_{\varrho}^2G^2(\widetilde{\theta})+
\sigma_{\eta}^2V^2(\varphi)\right]^{-1/2}d\widetilde{\theta}\nonumber\\[6pt]
+\sigma_{\eta}^4\,\frac{\zeta+2}{4}\,\frac{[(2\varphi^2-\sin^2\varphi)\cos\varphi-\varphi\sin\varphi]^2}{\varphi^2-\sin^2\varphi}\,.
\end{eqnarray}

\begin{figure}[ht]
\subfigure{\includegraphics[scale=0.5]{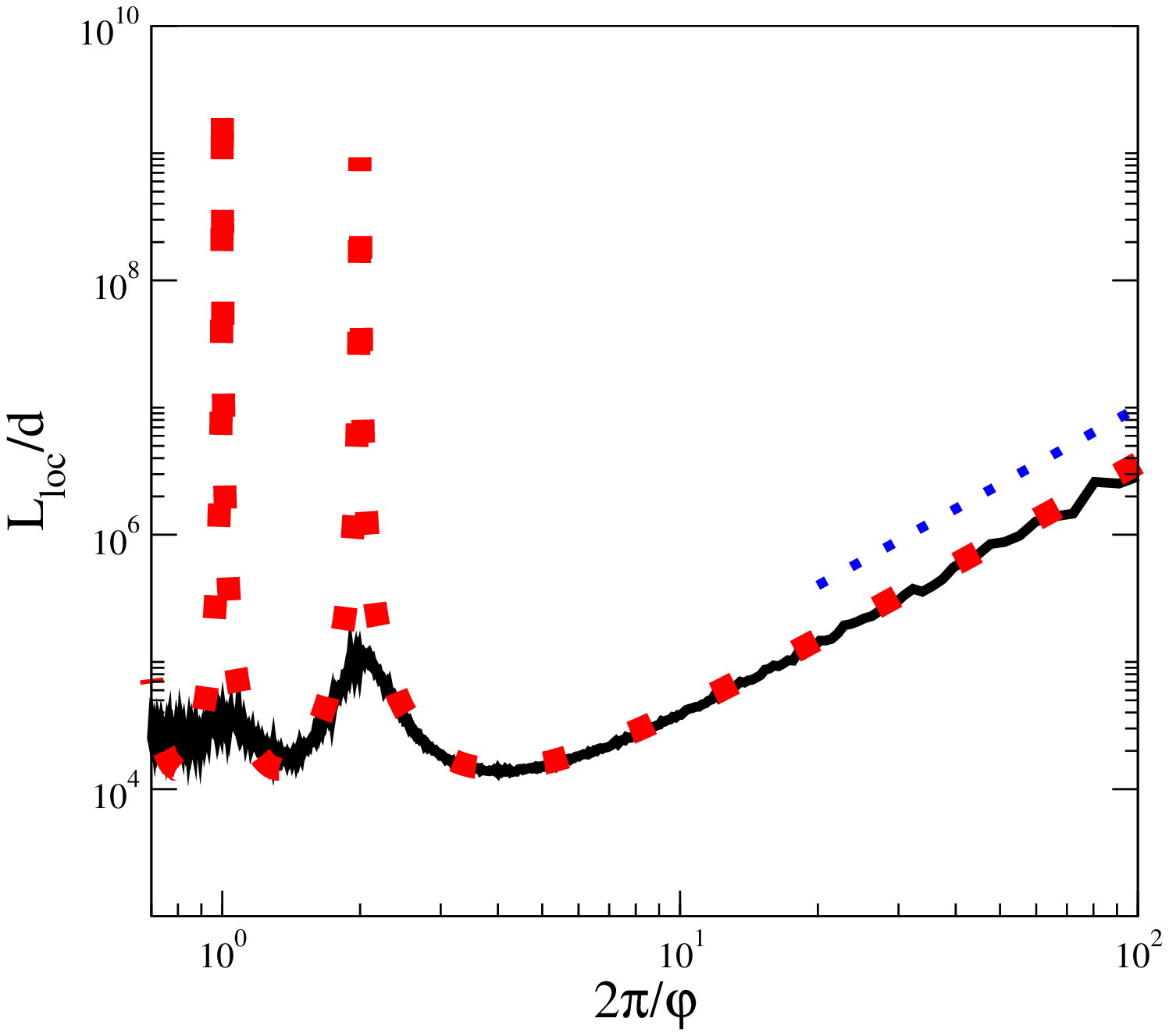}}
\subfigure{\includegraphics[scale=0.5]{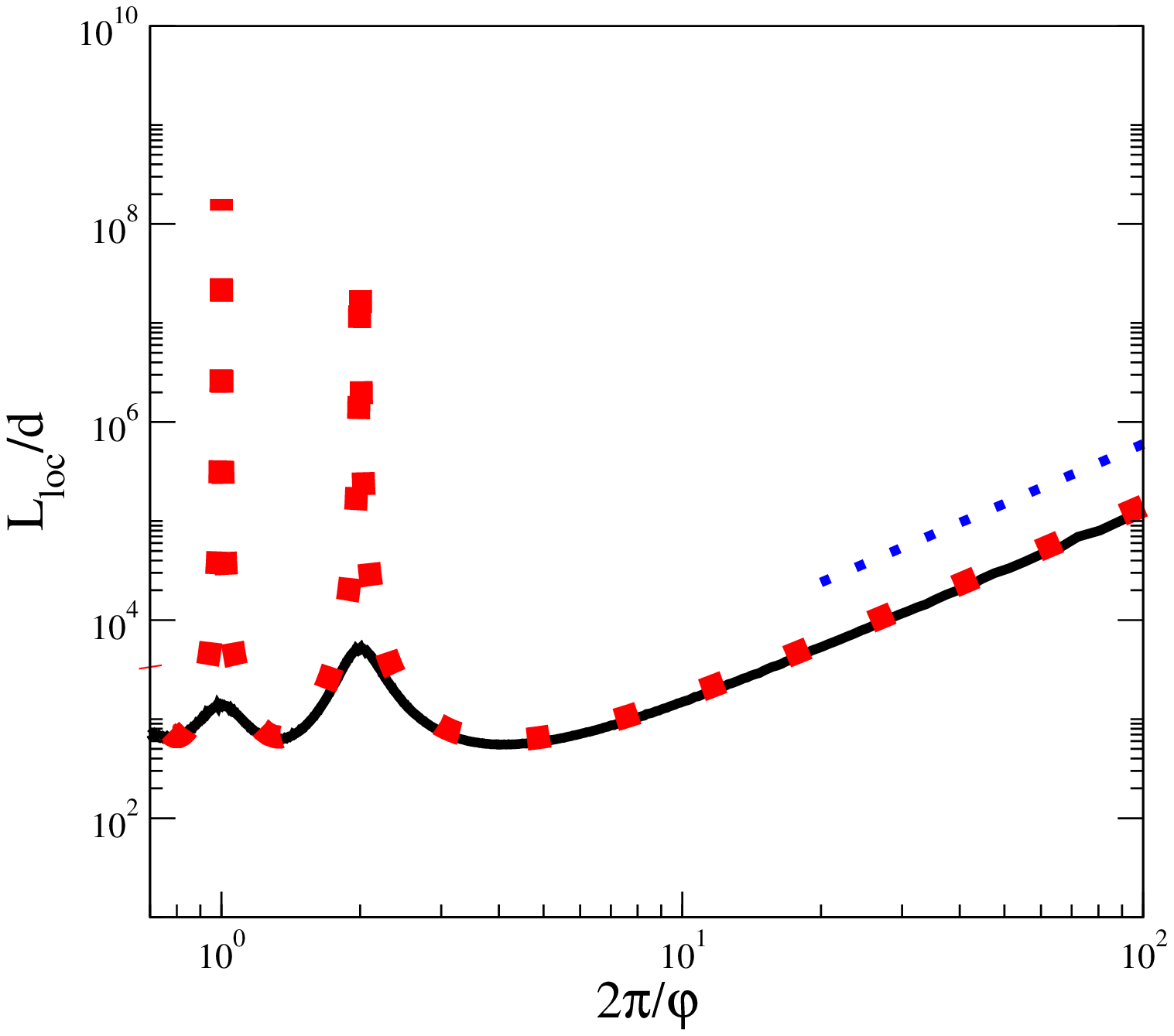}}
\caption{(Color online) Localization length versus the normalized wavelength $2\pi/\varphi=2\pi c/n_ad_a\omega$ for RH-LH matched quarter stack randomized by both types of disorder with $\sigma_{\varrho}^2=0.01$, $\sigma_{\eta}^2=0.00007$ (left), and $\sigma_{\varrho}^2=\sigma_{\eta}^2=0.003$ (right). The continuous curve presents numerical data for the structure length $N=10^7$ with an ensemble averaging performed over $10^3$ realizations of disorder. Large squares correspond to \eref{eq:CPD-LyapMixed} and small squares stand as a reference to $\lambda\propto\varphi^2$.}
\label{NJP-Fig05}
\end{figure}

\begin{figure}[t]
\begin{center}
\includegraphics[scale=0.6]{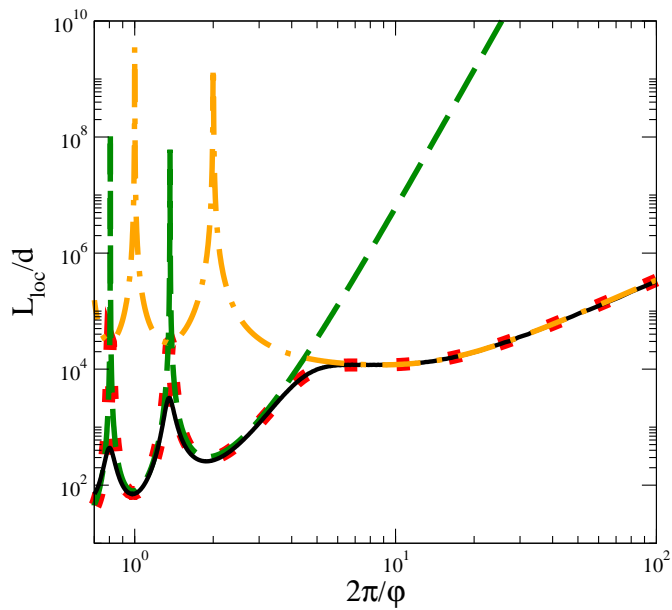}
\end{center}
\caption{(Color online) Same as in Figure \ref{NJP-Fig05} for $\sigma_{\varrho}^2=0.00003$ and $\sigma_{\eta}^2=0.02$. The continuous curve presents numerical data, full squares correspond to expression \eref{eq:CPD-LyapMixed}, its first and second terms are shown by dashed-dotted and dashed curves, respectively.}
\label{NJP-Fig06}
\end{figure}

From Figures \ref{NJP-Fig05} and \ref{NJP-Fig06} one can see that expression \eref{eq:CPD-LyapMixed} displays a very good agreement with the numerical data for the Lyapunov exponent obtained from the exact map \eref{eq:mapQP} for the mixed RH-LH matched quarter stack. It is important to emphasize that this expression is applicable for any ratio between the variances $\sigma_{\eta}^2$ and $\sigma_{\varrho}^2$. Also, one should note that it is valid within a wide frequency range restricted to the requirement \eref{eq:CPD-WeakDis} of a weak disorder only. Remarkably, in the absence of compositional disorder, $\sigma_{\eta}^2=0$, the Lyapunov exponent \eref{eq:CPD-LyapMixed} vanishes even if the positional disorder is non-zero, $\sigma_{\varrho}^2\neq0$. This fact is a direct consequence of the layer matching \eref{eq:Matching}, and is in the accordance with the previously obtained results \cite{IKM12,IM09}.

The first term in \eref{eq:CPD-LyapMixed} prevails over the second one and correctly describes the localization length when the positional disorder predominates, $\sigma_{\eta}^2\ll\sigma_{\varrho}^2$, or when both disorders are of the same order of magnitude, $\sigma_{\eta}^2\sim\sigma_{\varrho}^2$, see Figure \ref{NJP-Fig05}. The most important and interesting consequence of expression \eref{eq:CPD-LyapMixed} is that at any finite but sufficiently small strength $\sigma_{\eta}^2$ of the compositional disorder, its second term is negligible. For this case we arrive at simple, however, non-trivial result,
\begin{equation}\label{eq:CPD-LyapMixAs}
d/L_{loc}\equiv\lambda\approx\sigma_{\eta}^2\sin^2\varphi\qquad\mathrm{for}\quad\sigma_{\eta}^2\ll\sigma_{\varrho}^2,
\end{equation}
This asymptotics coincides with the Lyapunov exponent \eref{eq:CD-LyapHom} for the RH-RH matched quarter stack and provides the conventional quadratic frequency dependence, $\lambda\propto\sigma_{\eta}^2\omega^2$ when $\omega\to0$, see Figure \ref{NJP-Fig05}. One should emphasize that the result \eref{eq:CPD-LyapMixAs} is valid when the positional disorder predominates. However, due to the layer matching \eref{eq:Matching}, it does not contribute to the Lyapunov exponent. Thus, expression \eref{eq:CPD-LyapMixAs} turns out to be provided by small compositional disorder, instead of a large positional one. In spite of this fact, the presence of positional disorder plays the fundamental role, changing the abnormal octal frequency dependence \eref{eq:CD-LyapMixedOmega}, $L_{loc}^{-1}\propto\sigma_{\eta}^{4}\omega^{8}$, into the convetional quadratic one, $L_{loc}^{-1}\propto\sigma_{\eta}^2\omega^2$, when $\omega\to0$.

The second term in \eref{eq:CPD-LyapMixed} coincides with the Lyapunov exponent \eref{eq:CD-LyapMixed} caused by the compositional disorder only. However, when the positional disorder is also present, this term can significantly contribute to the Lyapunov exponent if the compositional disorder prevails over the positional one, $\sigma_{\varrho}^2\ll\sigma_{\eta}^2$. At first glance, in this case one can neglect the terms with $\sigma_{\varrho}^2$ in the radicand of the first summand in \eref{eq:CPD-LyapMixed} and in the radicand of the normalization function \eref{eq:CPD-I-def}. In other words, one can suggest that the distribution function \eref{eq:CPD-RhoMix} is uniform \eref{eq:CPD-RhoMixUn}. However, a more careful analysis shows that this is not true at relatively low frequencies $\omega$, when the phase shift $\varphi$ is sufficiently small, $\varphi^6\sigma_{\eta}^4\ll\sigma_{\varrho}^2\ll\sigma_{\eta}^2$. For such frequencies the Lyapunov exponent again obeys the standard quadratic $\omega$-dependence originated from the first term in \eref{eq:CPD-LyapMixed} and is described by expression \eref{eq:CPD-LyapMixAs}. Nevertheless, when the condition $\sigma_{\varrho}^2\ll\sigma_{\eta}^2$ is met, the second term does contribute to the localization length in the regime of high and intermediate frequencies. This fact is clearly displayed in Figure \ref{NJP-Fig06}.

\section{Conclusion}

In this paper we have studied two 1D layered models which consist of both right-handed (RH) and left-handed (LH) materials for alternative $a\,$- and $b\,$-layers of thicknesses $d_a$ and $d_b$, respectively. The problem considered above is an analytical evaluation of the Lyapunov exponent whose inverse value is the localization length. As was found in a set of numerical studies \cite{Ao07,Ao10}, for the RH-LH model a quite unexpected Anderson-type localization emerges, that has led to an intensive discussion in literature. In previous papers \cite{TIM11,TIM12} we have shown that the origin of this abnormal localization can be associated with a highly non-homogenous distribution of phase of the wave propagating along the bilayer structure. Unexpectedly, it was discovered \cite{TIM11} that with weak fluctuations of dielectric constants $\varepsilon_{an}$, $\varepsilon_{bn}$ the Lyapunov exponent vanishes in the second order of perturbation theory. This fact has shed light on the origin of the observed non-conventional localization. On the other hand, it was understood that the analytical treatment of this localization is of a very difficult task since one has to develop a proper perturbation theory beyond the second-order approximation.

The problem of this unusual localization was rigorously solved with the use of a new method, and brief communication was published in \cite{TIM12}. Above, we have presented an analysis of the problem with many details that are important to understand our approach, together with a discussion concerning the mechanism of the observed phenomenon.

We have to emphasize that originally the effect of abnormal localization was observed in \cite{Ao07,Ao10} for a quite specific bilayer model for which without disorder the RH-LH array represents an ideal mixed stack. Specifically, it was assume that $\varepsilon_a=\mu_a=1$, $\varepsilon_b=\mu_b=-1$, hence the impedances are trivially equal, $Z_a=Z_b=1$. The additional assumption was that the layer thicknesses are also equal, $d_a=d_b$. In the present study we consider a more general model of the so-called \emph{matched quarter stack} for which the condition of equal optical pass lengths, $n_ad_a=|n_b|d_b $, and the equality of the impedances, $Z_a=Z_b$, are independent of each other. Thus, our results can be applied to a more general class of physical systems.

In our approach we use the reduction of the underlying problem of a quantum localization to the study of the classical Hamiltonian map of a quite transparent structure. In this way the evaluation of the localization length can be performed in terms of local instability of trajectories in the phase space of the corresponding classical model. Then, with the transformation to radius-angle variables in the phase space, the equations allowing to evaluate the Lyapunov exponent take a quite simple form, and the problem is reduced to the analytical treatment of the angle distribution which emerges after a sufficiently long time of iteration of the classical map. It should be noted that the angle in this classical map is nothing but the phase of wave function in the quantum problem. Therefore, even with the reduction of the quantum model to the classical one the physical correspondence between two representations remains transparent.

Our approach allows one to resolve the problem of the Anderson localization for bilayer RH-LH arrays with the matched layers, and derive the localization length. As we show, the peculiarity of this model is entirely due to the zero value of the unperturbed Bloch phase that happens when the unperturbed impedances are equal. We rigorously derive the expression for the Lyapunov exponent which appears to be defined by the fourth order of perturbation theory, and which is valid for {\it any} value of frequency $\omega$. Our results prove that for small $\omega$ the localization length is enormously  large, $L_{loc}\propto\sigma_{\eta}^{-4}\omega^{-8}$. The generalization of our result is due to the $\widetilde{\theta}$-map \eref{eq:CD-mapTheta}, according to which the same dependence for $L_{loc}$ is expected to occur when the effective unperturbed refractive index $\overline{n}=(n_ad_a-|n_b|d_b)/(d_a+d_b)$, see \eref{eq:DR-Matching}, is sufficiently small, $\overline{n}\ll\sigma_{\eta}^2$. Thus, the obtained results show that the non-conventional localization observed numerically in Refs.\cite{Ao07,Ao10,Ao10a,Mo10} is very fragile with respect to the choice of model parameters, and can be hardly observed experimentally.

It is interesting that the discussed effect of anomalous localization can be considered as a kind of strongly correlated disorder, one part of which is embedded into the randomized phase shifts $\varphi_{an}$, $\varphi_{bn}$ and the other is absorbed by the impedances $Z_{an}$, $Z_{bn}$. As one can see, the compositional disorder, even if it is of a white-noise type, results in strong correlations between fluctuations of phase shifts and impedances. To the best of our knowledge, this phenomenon of a natural emergence of correlated disorder from the white-noise disorder was never discussed in literature.

A special attention in our study was paid to the model with two kinds of disorder. Specifically, we have considered the case when in addition to the compositional disorder with the fluctuating values of dielectric permittivities $\varepsilon_{an}$, $\varepsilon_{bn}$, the positional disorder is also included (therefore, the thicknesses $d_{an}$ and/or $d_{bn}$ also slightly fluctuate). It is extremely important that the positional disorder randomly affects the phase shifts only and do not contribute to the impedances. As one can see, the positional disorder plays a fundamental role for the destroy of strong correlations between the disordered phase shifts and impedances, therefore, for the recovering of the conventional dependence $L^{-1}_{loc}\propto\sigma_{\eta}^2\omega^2$ when $\omega\to0$.

The positional disorder in both thicknesses $d_{an}$ and $d_{bn}$ has been earlier analysed in \cite{IM09}. Specifically, the general case of correlated disorder was analytically studied by assuming any kind of statistical correlations in thicknesses of $a\,$- and $b\,$-layers, as well as the inter-correlations between the fluctuations of two thicknesses. It was shown that for any ratio between $d_a$ and $d_b$ the localization length is governed by the unique expression, no matter whether the structure represents RH-RH or RH-LH array. The main result is that for such structures the phase distribution is flat, therefore, the analytical expressions for the Lyapunov exponent can be obtained within the standard perturbation theory. The same statement stems from the analysis of RH-RH and RH-LH arrays with fluctuating refractive indices \cite{IMT10}, however, only when $n_ad_a\neq |n_b|d_b$ in the latter case. These results also indicate that the abnormal localization emerges in a very specific situation, and not only due to the inclusion of left-handed materials into the structure.

Finally, we would like to note that the effect of vanishing of the unperturbed Bloch phase shift $\gamma$ in the $\theta$-map \eref{eq:CD-mapTheta}, is a typical effect occurring at band edges in the tight-binding Anderson model as well as in the Kronig-Penney models. Indeed, in these models after one period of perturbation the unperturbed Bloch phase $\gamma$ vanishes when approaching the band edges. This results in a very specific non-homogeneous distribution of the wave phase, therefore, to a non-standard expression for the Lyapunov exponent. However, in the considered model of the mixed RH-LH matched quarter stack, the peculiarity is that vanishing Block phase emerges independently of the value of frequency $\omega$. This is a principal difference as compared with the localization occurring in the Anderson and Kronig-Penney models for which the zero Bloch phase emerges at band edges, therefore, only for specific values of frequency.

\ack
We gratefully acknowledge the support of the SEP-CONACYT (M\'exico) grant CB-2011-01-166382, as well as the support from VIEP grant EXC08-G of the BUAP (M\'exico).



\end{document}